\documentclass[english,aps,preprint]{paper}
\usepackage[LGR,T1]{fontenc}
\usepackage[latin9]{inputenc}
\usepackage{geometry}
\usepackage{color}
\usepackage{float}
\usepackage{units}
\usepackage{textcomp}
\usepackage{amstext}
\usepackage{amssymb}
\usepackage{graphicx}
\usepackage{setspace}
\usepackage{esint}
\makeatletter

\DeclareRobustCommand{\greektext}{%
  \fontencoding{LGR}\selectfont\def\encodingdefault{LGR}}
\DeclareRobustCommand{\textgreek}[1]{\leavevmode{\greektext #1}}
\DeclareFontEncoding{LGR}{}{}
\DeclareTextSymbol{\~}{LGR}{126}
\newcommand{\lyxmathsym}[1]{\ifmmode\begingroup\def\b@ld{bold}
  \text{\ifx\math@version\b@ld\bfseries\fi#1}\endgroup\else#1\fi}

\makeatother

\usepackage{babel}
\begin{document}

\title{On the Zitterbewegung Transient Regime in a Coarse-Grained Space-Time}

\author{J. Weberszpil$^{1}$ J. A. Helayël-Neto$^{2}$}

\maketitle
$^{1}$josewebe@ufrrj.br

Universidade Federal Rural do Rio de Janeiro, UFRRJ-IM/DTL\\
 Av. Governador Roberto Silveira s/n- Nova Iguaçú, Rio de Janeiro,
Brasil

$^{2}$helayel@cbpf.br

Centro Brasileiro de Pesquisas Físicas-CBPF-Rua Dr Xavier Sigaud 150,\\
 22290-180, Rio de Janeiro-RJ, Brasil. 
\begin{abstract}
In the present contribution, by studying a fractional version of Dirac's
equation for the electron, we show that the phenomenon of Zitterbewegung
in a coarse-grained medium exhibits a transient oscillatory behavior,
rather than a purely oscillatory regime, as it occurs in the integer
case, $\alpha=1$. Our result suggests that, in such systems, the
Zitterbewegung-type term related to a trembling motion of a quasiparticle
is tamed by its complex interactions with other particles and the
medium. This can justify the difficulties in the observation of this
interesting phenomenon. The possibility that the Zitterbewegung be
accompanied by a damping factor supports the viewpoint of particle
substructures in Quantum Mechanics.
\end{abstract}

\section{introduction}

Dirac's equation unifies both Quantum Mechanics and Special Relativity
by providing a relativistic description of the electron's spin; it
predicts the existence of antimatter and is able to reproduce accurately
the spectrum of the hydrogen atom. It also embodies the 'Zitterbewegung'
(ZB) effect as an unexpected quivering motion of a free relativistic
quantum particle, like the electron, for instance. This name was coined
by Schrödinger, who first observed that, in describing relativistic
electrons by the Dirac's equation, the components of the relativistic
quadri-velocity do not commute with the free-electron Hamiltonian,
with the consequence that the electron's velocity is not a constant
of the motion even in the absence of external fields. Such an effect
must be of a quantum nature, as it does not obey Newton's laws of
classical dynamics. Schrödinger calculated the resulting time dependence
of the electron's velocity and position, concluding that, in addition
to its classical motion, the electron experiences very fast periodic
oscillations \cite{Wlodek}.

One of the motivations for analyzing ZB-models is to describe the
spin of the electron, $S$, and its magnetic moment, $\lyxmathsym{\textgreek{m}}$,
as generated by a local circulation of mass and charge. Experiments
indicate the possibility for an internal structure for the electron,
considering it as an extended object (wavelength $10^{-13}$m $<\delta<$
$10^{-16}$m, where $\delta$ is the supposed dimension for the electron
\cite{Hestenes- Modeling}).\textbf{ }Another motivation is the possible
existence of a non-vanishing electric dipole moment for the electron
due to the separation between the center of mass, which is related
to the Foldy-Wouthuysen position operator, and the center of charge,
which corresponds to Dirac's position operator, x \cite{Chiral oscillations-Campinas}.

The trembling electron's motion should occur also in crystalline solids,
in classical systems, in macroscopic sonic crystals, in materials
like bi- and mono-layer graphene and nanotubes; it can also be predicted
to occur in the presence of an external applied magnetic field. In
spite of the great interest in the phenomenon of ZB, its physical
origin has remained mysterious. It was recognized that the ZB in vacuum
is due to an interference of states corresponding to positive and
negative electron energies \cite{Wlodek}.

Experimental difficulties to observe the ZB in vacuum are considerable
because the predicted frequency of the trembling is very high: $\hbar\lyxmathsym{\textgreek{w}}_{Z}=2m_{0}c^{2}\simeq1MeV$,
and its amplitude is very small: $\lyxmathsym{\textgreek{l}}_{c}=\hbar/m_{0}c\simeq3.86\times10^{\lyxmathsym{\textminus}3}\mathring{A}$
\cite{Wlodek 2}, and they are not accessible to current experimental
techniques.

Fractional Calculus (FC) is one of the generalizations of classical
calculus. It provides a redefinition of mathematical tools and it
seems very useful to deal with anomalous \cite{klafter,klafter2,klafter3,9,scalas,Hilfer2}
and frictional systems. Several applications of FC may be found in
the literature \cite{stanisvasky,Metzler,Glockle}. Presently, areas
such as field theory and gravitational models demand new conceptions
and approaches which might allow us to understand new systems and
could help in extending well-known results.

In this work, we investigate the fractional coarse-grained aspects
of the electron's ZB in a coarse-grained medium. Here, we take the
viewpoint that the fractionality associated to the complex interactions
may be responsible for a damping of the ZB oscillations that, in turn,
justify the difficulties in their experimental observations. This
claim highlights that our motivation to adopt the FC is more physical
than a simple mathematical extension. This becomes more explicit once
we argue that the possible justification for the difficulties in the
experimental measurements are originated from the complexity of the
interactions of the electrons, considered as a pseudo-particle ''dressed''
by the interactions and the medium. Here, we look at the dynamical
system as an open system that can interact with the environment and
we argue that FC can be an important tool to study open classical
and quantum systems. \cite{Tarasov 2012-Fract Oscillator}.

Our efforts to justify the use of FC in fractional systems with dissipative
systems and the relationship with complex systems, coarse-grained
medium, limit scale energy for the interactions and and non-integer
dimensions may be found in the papers of \cite{Aspects-nosso,Cresus-Jose-Helayel-caos,Nosso  G-Factor,Sympletic,Gianluca - PRL 2010}.

For problems related to the quantization of field theories, the reader
could consult \cite{Cresus e Everton,Kleinert,baleanu,goldfain}.
Also, in connection with our work, a fractional Riemann\textendash{}Liouville
Zeeman effect and an attempt to implement gauge invariance in fractional
field theories and an angular momentum algebra proposed with the Riemann-Liouville
formalism are reported in the paper of \cite{Fract Field Herrmann}.
Low-energy nuclear excitations have been studied in connection with
a fractional symmetric rigid rotor in order to calculate barionic
excitations \cite{Rigid Rotor-Herrmann}.

Here, we claim that the use of an approach to FC, based on a sequential
form of the modified Riemann-Liouville (MRL) fractional calculus \cite{Jumarie1},
is more appropriate to describe the dynamics associated with field
theory and particle physics in the space of nondifferentiable solution
functions, or in the scenario of a coarse-grained space-time.

Some backgrounds on coarse-grained media, references on fractal space-time
and the efforts to build up a solid foundation for the construction
of a geometry and field theory in fractional spaces may be found in
\cite{Aspects-nosso,Cresus-Jose-Helayel-caos,Nosso  G-Factor,Sympletic,Gianluca - PRL 2010}
and references therein.

Here, to achieve our goal, we pursue an investigation of the fractional
Dirac's equation for the electron, looking at dynamical systems as
open interacting structures. We show that the ZB phenomenon in a coarse-grained
medium exhibits a transient oscillatory behavior, rather than a purely
oscillatory regime, as it occurs in the integer case, $\alpha=1$.
Our result suggests that, in such systems, the ZB-type term related
to a trembling motion of the quasiparticle is tamed by the complex
interactions of the quasiparticles with other particles and the medium.
This may be the argument to justify the difficulties in the observation
of this rather interesting phenomenon. The possibility that the ZB
be accompanied by a damping supports the viewpoint of a particle substructure
in Quantum Mechanics.

Our paper is outlined as follows: In Section 2, we consider some relevant
remarks on the fractional derivative. In section 3, we present the
fractional approach to Zitterbewegung and finally, in Section 4, our
Discussion and Conclusions are cast.

\section{Some Remarks on the Fractional Derivative}

In the sequel, we adopt an alternative approach by considering a fractional
coarse-grained space-time rather than a fractional space of functions,
meaning that neither the space nor the time are infinitely thin, but
they rather exhibit ''thickness''. As the use of certain calculation
rules is essential to our approach, we briefly comment on this point,
before presenting these rules.

The Riemann-Liouville and Caputo approaches to FC are well-known and
have their rules rigorously proved, as the reader may find in the
standard textbooks \cite{Podlubny,Kilbas-Srivastava-Trujillo,Samko-Kilbas-Marichev,Miller-Ross,Ross}.\textcolor{red}{{}
}These well-tested definitions for fractional derivatives, referred
to as Riemann-Liouville's and Caputo's, have been frequently used
for several applications. In spite of their efficacy, they have some
dangerous pitfalls. For this reason, an interesting definition for
fractional derivative \cite{Jumarie- Acta Sin,Jumarie 2013}, the
so-called modified Riemann-Liouville (MRL) fractional derivative,
has been proposed which is less restrictive than other definitions.
Its basic expression is as follows: 
\begin{eqnarray}
D^{\alpha}f(x) & = & {\displaystyle \lim_{x\rightarrow0}\, h^{-\alpha}\sum_{k=0}^{\infty}{\alpha \choose k}\, f\left(x+(\alpha\,-\, k)h\right)}=\nonumber \\
 & = & \frac{{1}}{\Gamma(1-\alpha)}\frac{d}{{dx}}\,\int_{0}^{x}(x\,-\, t)^{-\alpha}\left(f(t)-f(0)\right)dt;\\
 & 0<\alpha<1.\nonumber 
\end{eqnarray}

But, by strictly referring to the context of the modified Riemann-Liouville
(MRL) formalism, to our mind it seems worthy to notice that the chain
rule, as well as the Leibniz product rule, had their mathematical
validity proven only recently, in view of the formal proof for the
Taylor expansion \cite{Demonst Taylor}. To be more precise, in the
MRL approach, the fractional Taylor expansion is the mathematical
basis for the validation of the chain rule and also for the Leibniz
product rule. We also emphasize that these rules can then be viewed
as good approximations. Then, we point out that the Leibniz rule used
here is a good approximation that comes from the first two terms of
the fractional Taylor series development, which holds only for nondifferentiable
functions \cite{Jumarie- Acta Sin,Jumarie 2013} and are as good and
approximated as the classical integer one. The mentioned rules are
quite similar to their counterparts in the integer-order calculus.

Following the MRL definition, we find that derivative of a constant
is zero, and second, we can use it for differentiable and non-differentiable
functions as well. For further details on MRL formalism, the reader
can follow \cite{Nosso  G-Factor,Jumarie- Acta Sin,Jumarie 2013,Livro Jumarie},
which contain all the basic for the formulation of a fractional differential
geometry in coarse-grained space, and refers to an extensive use of
coarse-grained phenomenon.

Here, another important comment is worthy and concerns certain definitions
referred to as local fractional derivatives, like the ones introduced
by \cite{Kolwankar1,Kolwankar 2,Kolwankar 3}, with several works
related to this approach; we also quote the papers of refs. \cite{calculus of local fractional derivatives,On the local fractional derivative,Carpintieri}
and the approaches with the Hausdorff derivative and with the so-called
fractal derivative \cite{Hausdorff or fractal derivative,comparative-Fractal -Fractional}.
All of the mentioned treatments seem to be applicable to power-law
phenomena. For some alternative definitions of fractional derivatives,
see \cite{Kobelev} and the interesting work \cite{A new Definition}.
Some relevant comments and remarks on the similarities between different
approaches or even on fractional q-calculus may be found in \cite{Metzler,Nosso  G-Factor,comparative-Fractal -Fractional,Caceres 2004,Bologna-Grigolini-1999,Richard Herrmann}.

In a recent article \cite{Letter -Weberszpil}, one of the authors
show that the Leibniz rule for fractional derivatives in a coarse-grained
medium treated as a Hölder-space yields a modified chain rule. The
latter can be safely applied, in agreement with alternative versions
of fractional calculus, in these classes of spaces or even in the
local versions of FC.

We could revisit the ZB phenomena by applying the Balankin's \cite{Balankin-Map,Balankin- Flow}
approach based on mapping to a fractional continuum \cite{Tarasov-Continuum}
and by adopting a local version of fractional called Hausdorff derivative.
Doing so, an expansion of the fractional Newton binomial, that appears
as a pre-factor in the derivative of \cite{Balankin-Map,Balankin- Flow},
can be shown to lead to a q-derivative \cite{Borges2004} as a lower-order
term \cite{Connections-Matheus}. The resulting equations would be
non-linear due to a local factor, with integer-order derivatives acting
on a continuum Euclidean space of fractional metric as a result of
the mapping \cite{Balankin-Map}. Alternatively, by applying the approaches
found in \cite{Kolwankar1,Kolwankar 2,Kolwankar 3,A new Definition,Adda-Cresson},
one could attain results close to the ones obtained here. Using another
definition of the local fractional derivative \cite{Xue-Feng Yang-Local Fractional},
which seams to be actually based on the MRL approach, but appear in
the form of a limit operation and can be seen by considering the first
two terms of a truncated fractional Taylor's series, would manifest
a global pre-factor, identical to that of the MRL formalism. This
global factor, $\Gamma(\alpha+1),$ could be considered as a statistical
average, that can be seen as a Mellin transform of exponential factors
$e^{-x},$ weighted by some statistical probability factor $x^{-\alpha}:$
$\Gamma(\alpha+1)=\int_{0}^{\infty}e^{-x}(x)^{-\alpha}dx$. We can
speculate that this could be interpreted in such a way to attribute
to each spatial point, $x$, a probability $(x)^{-\alpha}$ to influence
the system in a range $e^{-x}.$ The use of the global factor would
yield similar equations equations to those here described.

The connections above indicate that the MRL approach is consistent
with a local approach with a global pre-factor, when the fractional
Leibniz rule or the chain rule are used.

Now that we have set up these fundamental considerations, we are ready
to carry out the calculations of main interest in our paper.

\section{Fractional Approach to the Zitterbewegung}

In recent work \cite{Nosso  G-Factor}, we have built up a fractional
Dirac's equation in a coarse-grained scenario by taking into account
a fractional Weyl's equation, a fractional angular momentum algebra,
by introducing a mass parameter and imposing that the equations be
compatible with the fractional energy-momentum relation.

By extending the concept of helicity to include it in a fractional
scenario, we write down the left- and the right-handed Weyl's equations
from first principles in this extended framework \cite{Nosso  G-Factor}.
Next, by coupling the different fractional Weyl sectors by means of
a mass parameter, we arrive at the fractional version of Dirac's equation
as

\begin{equation}
(i\hbar^{\alpha}\gamma^{\mu}\partial_{\mu}^{\alpha}-{\bf 1}m^{\alpha}c^{\alpha})\Psi_{\alpha}=0.\label{eq:Fractional Dirac}
\end{equation}
Where 
\begin{equation}
\partial_{\mu}^{(\alpha)}=(\frac{M_{t,\alpha}}{c^{\alpha}}\frac{\partial^{\alpha}}{\partial t^{\alpha}};M_{x,\alpha}\frac{\partial^{\alpha}}{\partial x^{\alpha}}).
\end{equation}
With 
\begin{equation}
\widehat{p}^{\alpha}=-i\left(\hbar\right)^{\alpha}M_{x,\alpha}\frac{\partial^{\alpha}}{\partial x^{\alpha}},\label{eq:Oper p_alpha}
\end{equation}
we have the fractional commutation relation $[\hat{x}_{\imath}^{\alpha},\hat{p}_{j}^{\alpha}]=\imath\Gamma(\alpha+1)\hbar^{\alpha}M_{\alpha}\delta_{\imath j}I.$

The fractional Dirac Hamiltonian is then \cite{Nosso  G-Factor} 
\begin{equation}
H^{\alpha}=c^{\alpha}\vec{\alpha}\cdot\vec{p}^{\alpha}+m^{\alpha}c^{2\alpha}\beta.\label{eq:Fractional Dirac Hamiltonian}
\end{equation}

Applying the operator eq.(\ref{eq:Fractional Dirac Hamiltonian})
to a state-vector $\psi_{\alpha},$ and using that $i\hbar^{\alpha}{}_{0}^{J}D_{t}^{\alpha}\psi_{\alpha}=H^{\alpha}\psi_{\alpha},$
yields

\begin{equation}
i\hbar^{\alpha}{}_{0}^{J}D_{t}^{\alpha}\psi_{\alpha}=(c^{\alpha}\vec{\alpha}\cdot\vec{p}^{\alpha}+m^{\alpha}c^{2\alpha}\beta)\psi_{\alpha}=H^{\alpha}\psi_{\alpha}.
\end{equation}

We now write that the state-vector $\psi_{\alpha}(\vec{r,t})$ as
below: 
\begin{equation}
\psi_{\alpha}(\vec{r,t})=U_{\alpha}(t,t_{0})\psi_{\alpha}(\vec{r,t}_{0}),
\end{equation}
where $U_{\alpha}(t,t_{0})$ is the fractional temporal evolution
operator to be determined.

This determination can be set as follows: 
\begin{equation}
i\hbar^{\alpha}{}_{0}^{J}D_{t}^{\alpha}U_{\alpha}(t,t_{0})\psi_{\alpha}(\vec{r,t}_{0})=H^{\alpha}U_{\alpha}(t,t_{0})\psi_{\alpha}(\vec{r,t}_{0}).
\end{equation}

So, the evolution equation may be written as 
\begin{equation}
i\hbar^{\alpha}{}_{0}^{J}D_{t}^{\alpha}U_{\alpha}(t,t_{0})=H^{\alpha}U_{\alpha}(t,t_{0}).\label{eq:evolution Equation}
\end{equation}

Noticing that $_{0}^{J}D_{t}^{\alpha}E_{\alpha}(\lambda x^{\alpha})=\lambda E_{\alpha}(\lambda x^{\alpha}),$we
can see that the solution to the eq. (\ref{eq:evolution Equation})
is the fractional evolution operator in a coarse-grained scenario,
that is given by $U_{\alpha}(t,t_{0})=E_{\alpha}(-\frac{i}{\hbar^{\alpha}}H^{\alpha}t^{\alpha})$.
Here, we have supposed that $D_{t}^{\alpha}H^{\alpha}=0,$ and $E_{\alpha}(x)$
is the one-parameter Mittag-Leffler function.

Going now into the Heisenberg representation, an operator $A_{\alpha}^{H}$ca
be written as

\begin{equation}
A_{\alpha}^{H}=U_{\alpha}^{+}(t,0)A_{\alpha}^{S}U_{\alpha}(t,0),
\end{equation}
where $A_{\alpha}^{S}$is the operator $A_{\alpha}$in the Schrödinger
representation.

In order to set up a fractional version of Heisenberg evolution equation,
we can use the fractional Leibniz rule in the MRL approach. The result
is 
\begin{equation}
_{0}^{J}D_{t}^{\alpha}A_{\alpha}^{H}(t)=-\frac{i}{\hbar^{\alpha}}[A_{\alpha}^{H},H^{\alpha}].
\end{equation}

We can now calculate the fractional evolution of the position as 
\begin{equation}
_{0}^{J}D_{t}^{\alpha}x_{\alpha}^{H}(t)=-\frac{i}{\hbar^{\alpha}}[x_{\alpha}^{H},H^{\alpha}].
\end{equation}

The commutation relations yield:

\begin{equation}
[H^{\alpha},x_{\alpha}^{H}],=[c^{\alpha}\vec{\alpha}\cdot\vec{p}^{\alpha}x_{\alpha}^{H}]=-c^{\alpha}\vec{\alpha}\imath\Gamma(\alpha+1)\hbar^{\alpha},
\end{equation}
so that the fractional evolution equation turns out to the be the
fractional Breit equation: 
\begin{equation}
_{0}^{J}D_{t}^{\alpha}x_{\alpha}^{H}(t)=\Gamma(\alpha+1)c^{\alpha}\vec{\alpha}.\label{eq:evol_x.vs.Alpha}
\end{equation}
As a next step, we proceed to calculate $_{0}^{J}D_{t}^{\alpha}\vec{\alpha}.$

Using that the Poisson bracket, 
\begin{equation}
\{H^{\alpha},\vec{\alpha}\}=2c^{\alpha}\vec{p}^{\alpha},
\end{equation}
and the fractional Heisenberg equation, we get that 
\begin{equation}
_{0}^{J}D_{t}^{\alpha}\vec{\alpha}=\frac{i}{\hbar^{\alpha}}(-2c^{\alpha}\vec{p}^{\alpha}+2H^{\alpha}\vec{\alpha}),
\end{equation}
or 
\begin{equation}
i\hbar^{\alpha}{}_{0}^{J}D_{t}^{\alpha}\vec{\alpha}=2H^{\alpha}\vec{\eta}_{\alpha},
\end{equation}
where $\vec{\eta}_{\alpha}$ is given by 
\begin{equation}
\vec{\eta}_{\alpha}\equiv\vec{\alpha}-c^{\alpha}(H^{\alpha})^{-1}\vec{p}^{\alpha}.
\end{equation}

Since $_{0}^{J}D_{t}^{\alpha}(H^{\alpha})^{-1}\vec{p}^{\alpha}=0$,
we can write that 
\begin{equation}
_{0}^{J}D_{t}^{\alpha}\vec{\alpha}={}_{0}^{J}D_{t}^{\alpha}\vec{\eta}_{\alpha}.
\end{equation}
Now, assuming that the conservation relations hold 
\begin{equation}
_{0}^{J}D_{t}^{\alpha}H^{\alpha}=_{0}^{J}D_{t}^{\alpha}\vec{p}^{\alpha}=0,
\end{equation}
and by verifying that $\{H^{\alpha},\vec{\eta}_{\alpha}\}=0$, then,
we can write the fractional differential equation for $\vec{\eta}_{\alpha}$
as 
\begin{equation}
_{0}^{J}D_{t}^{\alpha}\vec{\eta}_{\alpha}=-\frac{i}{\hbar^{\alpha}}2H_{\alpha}\vec{\eta}_{\alpha}.\label{eq:evol_Eta}
\end{equation}
The solution to the eq.(\ref{eq:evol_Eta}) above is 
\begin{equation}
\vec{\eta}_{\alpha}=\vec{\eta}_{\alpha}(0)E_{\alpha}(-\frac{i}{\hbar^{\alpha}}2H^{\alpha}t^{\alpha});
\end{equation}
which can be rewritten as 
\begin{equation}
\vec{\alpha}=c^{\alpha}(H^{\alpha})^{-1}\vec{p}^{\alpha}+\vec{\eta}_{\alpha}(0)E_{\alpha}(-\frac{i}{\hbar^{\alpha}}2H^{\alpha}t^{\alpha}).
\end{equation}

With the help of eq. (\ref{eq:evol_x.vs.Alpha}), we have 
\begin{equation}
_{0}^{J}D_{t}^{\alpha}x_{\alpha}^{H}(t)=\Gamma(\alpha+1)c^{2\alpha}(H^{\alpha})^{-1}\vec{p}^{\alpha}+\Gamma(\alpha+1)c^{2\alpha}\vec{\hbar^{\alpha}\eta}_{\alpha}(0)E_{\alpha}(-\frac{i}{\hbar^{\alpha}}2H^{\alpha}t^{\alpha}).
\end{equation}

The fractional integration of the equation above results as below:
\begin{eqnarray}
x_{\alpha}(t) & = & \Gamma(\alpha+1)c^{2\alpha}(H^{\alpha})^{-1}\vec{p}^{\alpha}t^{\alpha}+\nonumber \\
 & + & \frac{i}{2}\Gamma(\alpha+1)c^{2\alpha}(H^{\alpha})^{-1}\vec{\hbar^{\alpha}\eta}_{\alpha}(0)E_{\alpha}(-\frac{i}{\hbar^{\alpha}}2H^{\alpha}t^{\alpha})+\nonumber \\
 & - & \frac{i}{2}\Gamma(\alpha+1)c^{2\alpha}(H^{\alpha})^{-1}\hbar^{\alpha}\vec{\eta}_{\alpha}(0).\label{eq:Final}
\end{eqnarray}
It can readily be checked that, for $\alpha=1,$ the result is the
same as in the integer case \cite{Sidarth-Revisiting}.

Now, for the Mittag-Leffler function the relation below holds: 
\begin{equation}
E_{\alpha}(-ix)=E_{2\alpha}(-x^{2})-ixE_{2\alpha,1+\alpha}(-x^{2}),
\end{equation}
where $E_{\gamma,\delta}(x)$ is the two-parameter Mittag-Leffler
function.

So, the real part of eq.(\ref{eq:Final}) can be represented as depicted
in the plot of Fig.1.

\begin{figure}[H]
\includegraphics[scale=0.4]{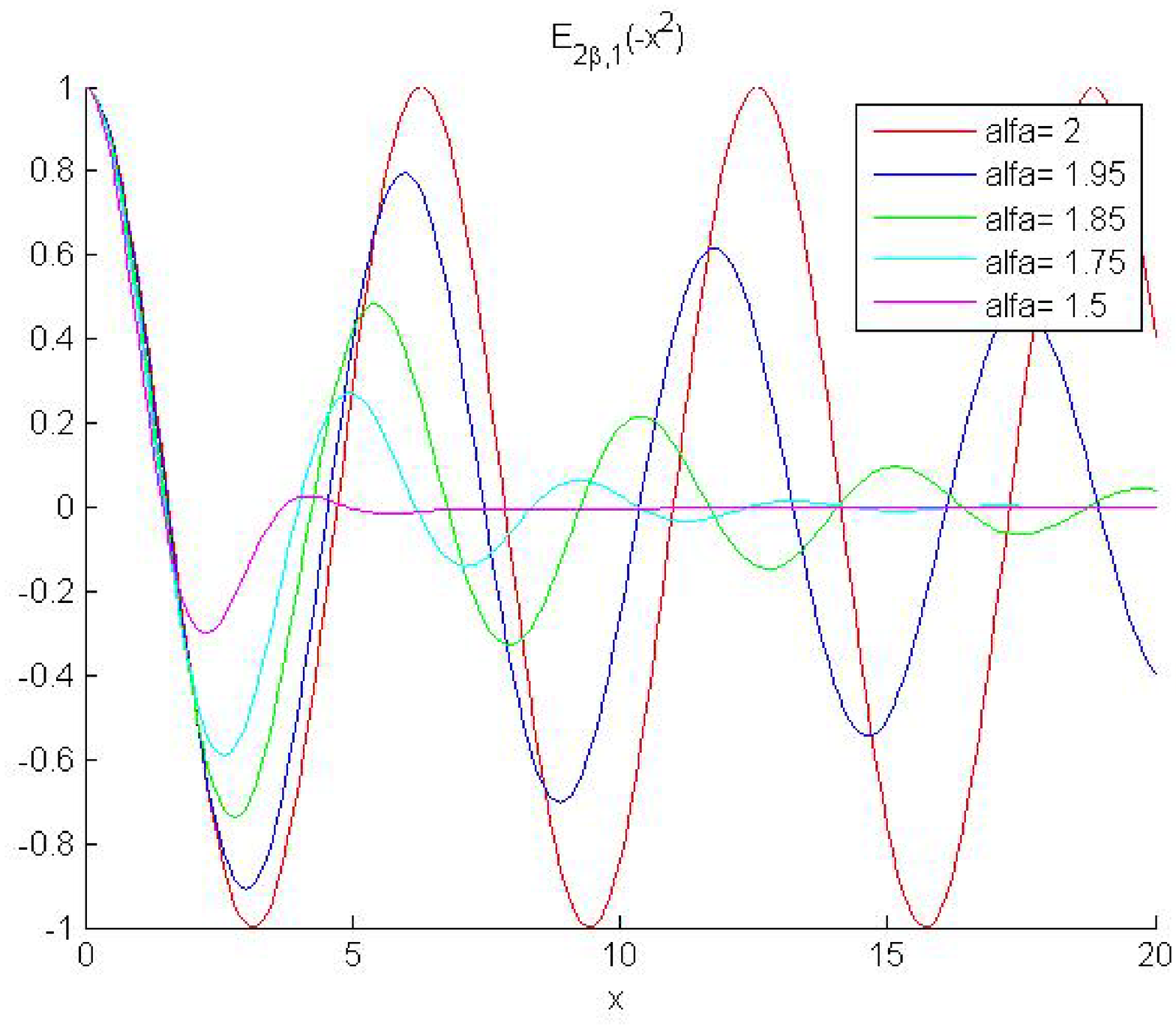}

Figure 1: ($x(t)=x_{0}E_{2\beta,1}(-\omega^{2}t^{2}).$ In the figure
$\alpha=2\beta$)
\end{figure}

\section{Discussion and Conclusions}

With the result cast in the previous Section, the evolution of the
particle's position can be determined, at every instant of time, by
a numerical simulation. It exhibits a transient oscillatory behavior
rather than a pure oscillatory regime, as in the integer case ($\alpha=1$).
This can justify the difficulties in the experimental observation
of the ZB phenomenon.

It is worthy to mention that quantum simulations of one-dimensional
Dirac's equations for free relativistic electrons carried out by Gerritsma
et al.\cite{Nature-Gerritsma} show that the ZB of an electron prepared
in the form of a Gaussian wave packet decays in time. This reproduces
the same behavior observed in the crystalline solids in Ref. \cite{Wlodek 2},
where the authors claim that the ZB represents the basic way of electron
propagation in a periodic potential and observe that the ZB of an
electron prepared in the form of a Gaussian wave packet decays in
time.

We can argue that the damping in the oscillations can be understood
by considering electrons as pseudo-particles, ''dressed'' by the
interactions and the medium and by looking at the dynamical system
as an open system that can interact with the environment. The complexity
of the interactions affect the value of the $\alpha-$fractional parameter.

Other important point to remark here is the possibility of establishing
connection between different formalisms, as we have pointed out in
Section 2. Indeed, by applying Balankin's approach with a version
of local fractional derivative, namely, the Hausdorff derivative,
and by considering an expansion of the fractional Newton binomial,
we are led to a q-derivative as a lower-order term. This may indicate
that the use of q-calculus in the context of non-additive entropy
is really justifiable, at least to the first order, in the realm of
complexity, fractals and multifractals, where power-law phenomena
take place. Efforts to better investigate this particular point are
in progress and we shall be reporting on that soon \cite{Connections-Matheus}.
Also, some higher-order terms in the fractional binomial expansion
could be included in order to yield better consistent theories. The
perspective for the construction of non-linear models, with integer
derivatives, in the context of complex systems is also a viable possibility,
by the mapping with the fractional continuum and the local fractional
Hausdorff derivative. Actually, non-linear fractional theories may
arise from the direct use of some versions of local fractional derivatives.

On the other hand, in the framework of Particle Physics, all massive
leptons, including their corresponding neutrinos \cite{ZB Neutrino}
should manifest the ZB effect. The confirmation of this claim in intimately
related to the experimental limitations and also to the quantum-mechanical
uncertain principle.

In fact, chirality corresponds to a property of fundamental importance
in the study of neutrino physics \cite{Bernardini EPJC,Bernardini PRD}.
The effects of chiral oscillations can be explained as a consequence
of the ZB phenomenon which emerges whenever solutions to Dirac's equation
are used to describe the space-time evolution of a wave packet of
massive particles like neutrinos \cite{Chiral oscillations-Campinas}.
In the work of Ref. \cite{Neutrino Chiral Oscillations-Leo-Rotelli},
there are suggestions that a $spin-\nicefrac{1}{2}$ particle produced
in a localized condition is subject to chiral oscillations remnant
from Zitterbewegung.

\textbf{\bigskip{}
}

\textbf{Acknowledgment:} The authors wish to express their gratitude
to FAPERJ-Rio de Janeiro and CNPq-Brazil for the partial financial
support. Prof. A. E. Bernardini is also acknowledged for helpful comments
and for pointing out pertinent references.\textbf{\bigskip{}
}

\end{document}